\documentclass[apjl]{emulateapj}
\usepackage{epstopdf}
\usepackage{natbib}
\usepackage{amsmath}
\usepackage{enumerate}
\begin{document}
\title{Density Jumps Near the Virial Radius of Galaxy Clusters}

\author{Anna Patej\altaffilmark{1} and Abraham Loeb\altaffilmark{2}}
\altaffiltext{1}{Department of Physics, Harvard University, 17 Oxford St., Cambridge, MA 02138}
\altaffiltext{2}{Harvard-Smithsonian Center for Astrophysics, 60 Garden St., Cambridge, MA 02138}

\email{Electronic Addresses: apatej@physics.harvard.edu, aloeb@cfa.harvard.edu}

\begin{abstract}
Recent simulations have indicated that the dark matter halos of galaxy clusters should feature steep density jumps near the virial radius. Since the member galaxies are expected to follow similar collisionless dynamics as the dark matter, the galaxy density profile should show such a feature as well. We examine the potential of current datasets to test this prediction by selecting cluster members for a sample of 56 low-redshift ($0.1 < z < 0.3$) galaxy clusters, constructing their projected number density profiles, and fitting them with two profiles, one with a steep density jump and one without. Additionally, we investigate the presence of a jump using a non-parametric spline approach. We find that some of these clusters show strong evidence for a model with a density jump. We discuss avenues for further analysis of the density jump with future datasets.
\end{abstract}

\defcitealias{diemer14}{DK14}
\defcitealias{rines13}{R13}

\section{Introduction}\label{s:intro}
Galaxy clusters contain a representative sample of the matter in the universe: the dominant constituent is dark matter, while the baryonic components include hot gas and the galaxies themselves \citep[for a review, see][]{voit05}. A longstanding analytical prediction for cosmological structure formation is the existence of a shock bounding the hot, gaseous intracluster medium and a corresponding jump in the dark matter profile, coincident with the virial radius of the galaxy cluster \citep[e.g.,][]{bertschinger85}. More recently, simulations by \citet{diemer14} (hereafter \citetalias{diemer14}) have reinforced the expectation that the dark matter halo itself should exhibit a sharp density steepening near the virial radius \citep[see also][]{adhikari14}. 

The member galaxies of a cluster are expected to trace the cluster's dark matter profile in the cluster outskirts, as they are subject to similar collisionless dynamics; accordingly, we may expect to see such a feature not just in the dark matter profile but also in the galaxy density profile. \citet{tully10} examined the distributions of the galaxies in the Coma and Virgo clusters and detected sharp density cut-offs at radii of 3 and 2 Mpc, respectively, from the cluster centers, which he identified with the caustics of second turnaround. These correspond roughly to the virial radii of these clusters \citep[e.g.,][]{kubo07,karachentsev14}. The analysis of \citet{trentham09} identified a similar feature in the galaxy group NGC 1023.

\citet{more15} mentioned that hints of such a jump may have been seen in other data sets. For instance, \citet{tully15} measured the second turnaround radius, which is associated with the aforementioned density jumps, of several groups and clusters. Additionally, \citet{rines13} (hereafter \citetalias{rines13}) presented spectroscopic velocities for cluster members in 58 galaxy clusters, from which they reconstructed the density profiles, which appear to show a deficit of galaxies with respect to an NFW profile \citep{nfw}. However, they noted that their results at large radii from the cluster center may be affected by having few spectroscopically observed cluster members in the exterior regions.

In this paper, we focus on the \citetalias{rines13} sample of clusters and use public data from the Sloan Digital Sky Survey (SDSS) \citep{york00} to select cluster members photometrically. Rather than use spectroscopic velocities, we construct the radial number density profiles of member galaxies and fit them with two functional forms to examine whether we can detect a feature consistent with the predicted density jump in any of the clusters using available data sets. Where applicable, we assume the standard $\Lambda$CDM cosmology with $\Omega_{\Lambda} = 0.73$ and $\Omega_{m}=0.27$, consistent with the parameters selected by \citetalias{diemer14} to enable direct comparison. To compare results across clusters we also use the measure $R_{\Delta}$, which is the radius within which the mean mass density $\bar{\rho} = \Delta\rho_b(z)$, where $\rho_b$ is a specified cosmological background density and $\Delta$ is the density contrast with respect to $\rho_b(z)$.  

\section{Data}\label{s:data}
\subsection{SDSS Catalogs}
To probe the existence of the density jump predicted by \citetalias{diemer14}, we focus our attention on the sample of 56 low-redshift ($0.1 < z < 0.3$) galaxy clusters from \citetalias{rines13} (of their original 58 clusters, we do not include MS0906/A750 in our samples, since these clusters are nearly coincident and hence it is difficult to make a clean selection of member galaxies for each). Since their cluster sample was selected from regions of sky covered by SDSS, we use publicly available data from the most recent Data Release 12 \citep{sdssdr12} to select cluster members. 

We obtain photometric catalogs from the SDSS SkyServer SQL server\footnote{http://skyserver.sdss.org/dr12/en/tools/search/sql.aspx}. For each cluster, we query galaxies within 1.5 degrees of the cluster center (which, from \citetalias{rines13}, is the X-ray center) in both right ascension (RA) and declination. We further restrict our query to sources in the `Galaxy' view with observed $r$-band magnitude between 14 and 22 and which have `clean' photometry as determined by the SDSS pipeline. For each galaxy, we obtain the dereddened `model' magnitudes (which we will use throughout the remainder of this work), corrected for Galactic extinction according to \citet{schlegel98} by the SDSS pipeline, and we also select photometric redshift estimates and associated redshift quality estimates from the `Photo-z' table.

\subsection{Cluster Member Selection}\label{s:clustermems}
We will test two cluster member selections. The first, which we will refer to as Selection A, is based solely on photometry, primarily comprising red cluster galaxies selected from their location on a color-magnitude diagram but also including some galaxies with sufficiently secure photometric redshift estimates that are blueward of the red sequence. The second, which we call Selection B, folds in the spectroscopy of \citetalias{rines13} to include additional cluster members and to reject some of the photometrically selected galaxies whose redshift estimates place them beyond the cluster. These selection procedures are outlined in more detail in the following sections.

\subsubsection{Red Sequence Cluster Member Selection}\label{s:rs}
Both of our methods of cluster member selection are based upon the target selection process of \citetalias{rines13}, which relied on the red sequence method of \citet{gladders00}. The red sequence refers to a linear feature in the color-magnitude diagram of galaxies in a cluster field that arises from the population of red, early-type galaxies that have been observed to comprise the majority of cluster members. By selecting galaxies within a range around this line, we can obtain a sample of cluster members. As in \citetalias{rines13}, we construct a diagram of $g-r$ vs $r$ and fix the slope of the red sequence line to $-0.04$. We then select the appropriate intercept by examining the distribution of galaxies within $5\;\mathrm{arcminutes}$ of the cluster center. We can also use the spectroscopic data publicly available from \citetalias{rines13}; by matching their spectroscopic catalogs to SDSS photometry, we can additionally calibrate our red sequence selection, as will be described in Section \ref{s:addcm}. 

\begin{figure*}
\begin{center}
\includegraphics[scale=0.25,trim = 0mm 10mm 20mm 10mm,clip]{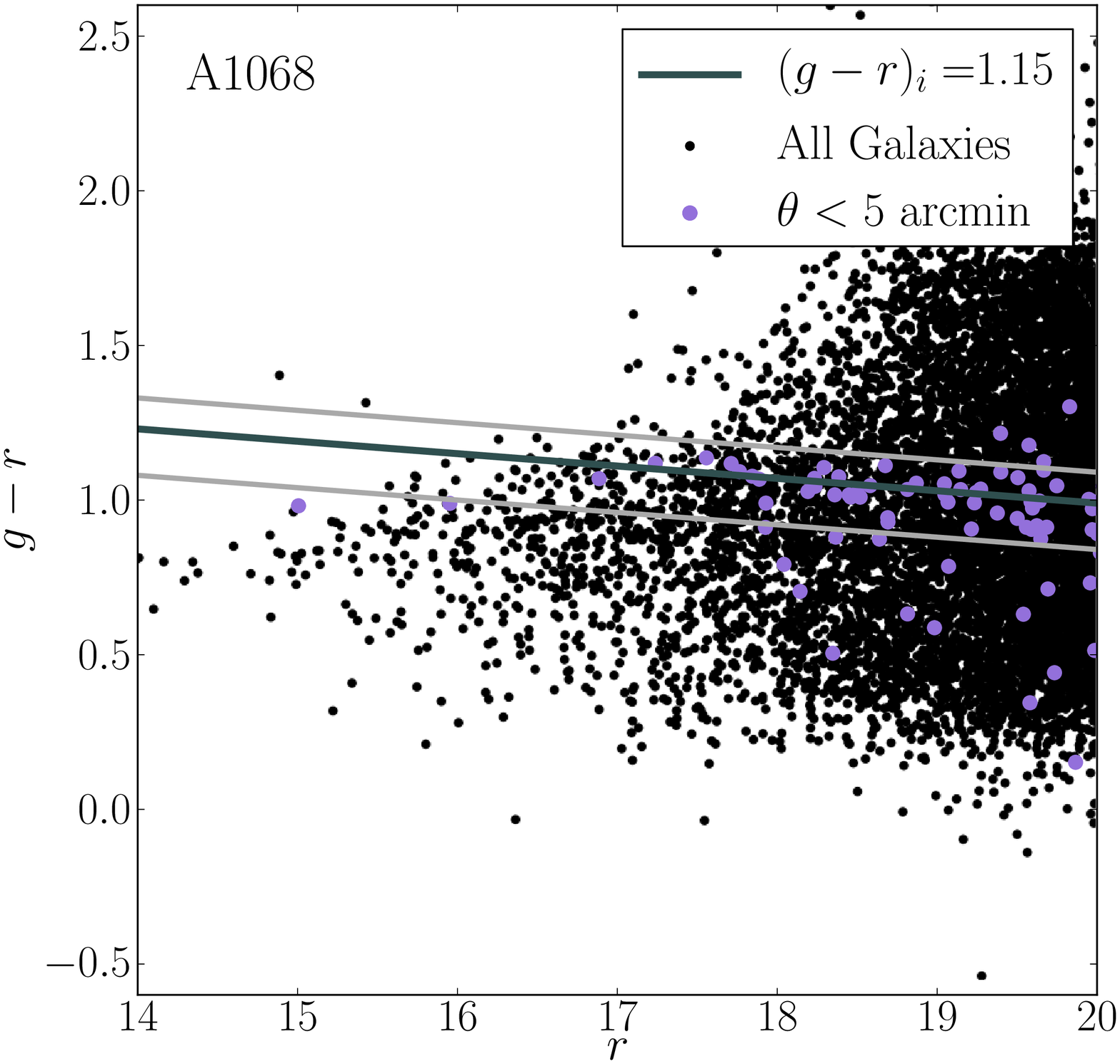}
\includegraphics[scale=0.25,trim = 0mm 10mm 20mm 10mm,clip]{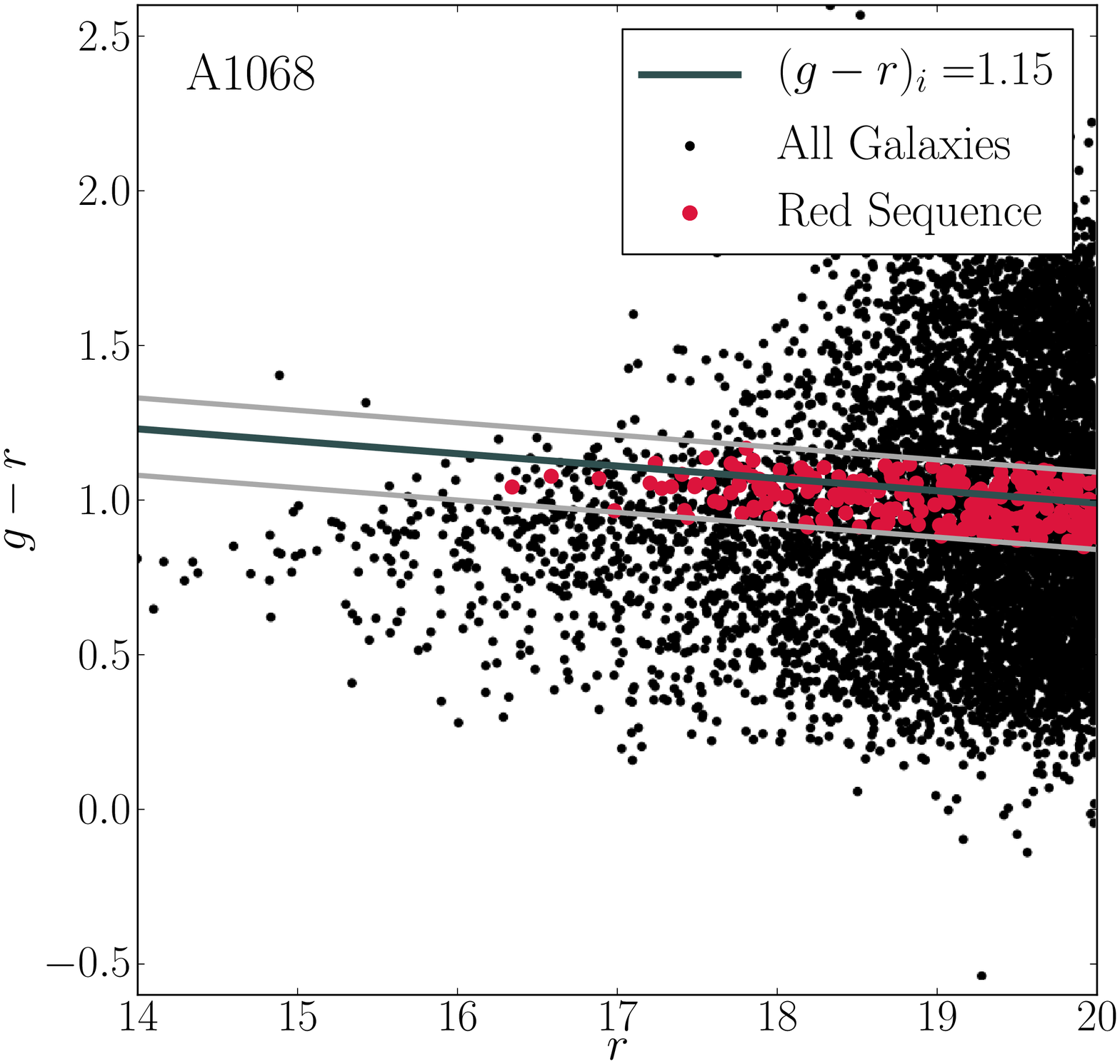}
\includegraphics[scale=0.25,trim = 0mm 10mm 20mm 10mm,clip]{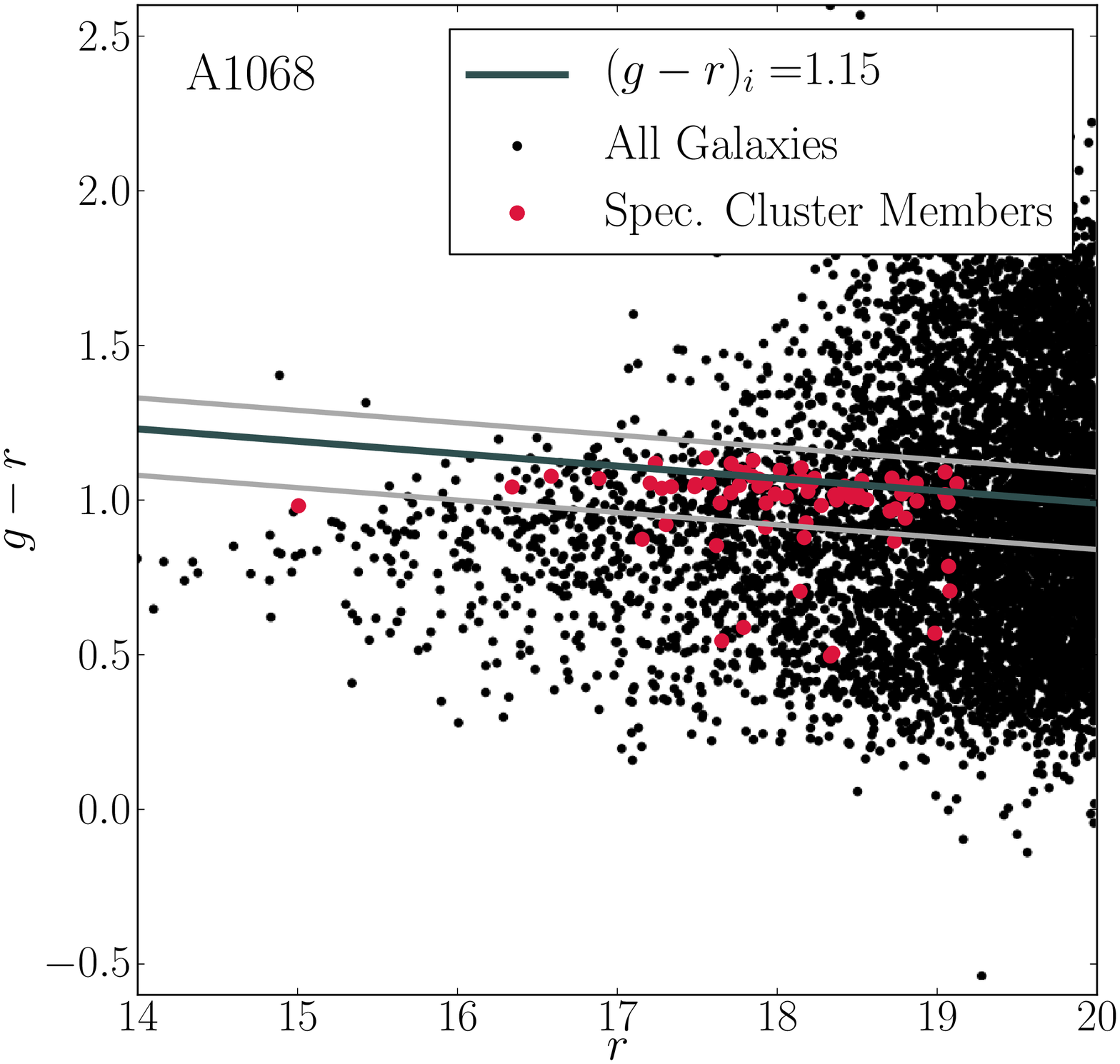}
\end{center}
\caption{The red sequence cluster member selection for example cluster A1068. \textit{Left}: in the color-magnitude diagram, we initially plot the location of galaxies within $5$ arcminutes of the cluster center, which should be dominated by cluster members, to provide an indication of the red sequence. \textit{Middle}: The galaxies selected by the red sequence method using the limits described in Section~\ref{s:rs}; we plot the galaxies thus selected that are within 2.5 $h^{-1}\mathrm{Mpc}$ of the cluster center. The labelled intercept $(g-r)_i$ is selected to be the value at $r=16$. \textit{Right:} The cluster members as selected via the spectroscopy of \citetalias{rines13}, which can be used to verify the red sequence selection.}\label{f:cm_rs}
\end{figure*}

Having thus identified the linear red sequence feature, we select galaxies that are within an offset of 0.1 magnitude above the red sequence line, and 0.15 magnitude below the red sequence line, as illustrated in Figure~\ref{f:cm_rs}. This is in contrast to the choice of $\pm0.3$ magnitude in \citetalias{rines13}, which was found to be much wider than the actual red sequence as indicated by the spectroscopy. The asymmetry of the limits above and below the red sequence is motivated by the results of Section 5.2 and Figure 16 of \citetalias{rines13}, which suggest that there are fewer cluster members above the red sequence than below it, as confirmed by the righthand panel of Figure~\ref{f:cm_rs}, which shows the distribution of spectroscopically confirmed members. 

\subsubsection{Contribution of Non-Red Sequence Galaxies}\label{s:blue}
Since the red sequence method selects a specific population of galaxies, it is worth examining whether the exclusion of other types of galaxies in the cluster affects our result. The analysis of \citet{dressler80}, which examined the galactic populations of over 50 galaxy clusters, found that blue, late-type galaxies comprise a small fraction of the populations of rich galaxy clusters, although this fraction increases as the density of galaxies decreases. This is supported by the spectroscopic data provided by \citetalias{rines13}, who find that within the virial radius of the cluster the fraction of blue galaxies is $\lesssim10\%$, while over the entire radial range for which they obtained spectra, the fraction is $\lesssim15\%$. 

Since we have photometric redshift estimates, we use these to also select cluster members that are blueward of the red sequence. We restrict our attention to galaxies that have secure photometric redshifts, as indicated by the `photoErrorClass' and `zErr' tags; the former we select to be equal to 1, and the latter we restrict to be less than 0.03. We then mark as additional cluster members galaxies that were not originally selected via the red sequence method and whose photometric redshift estimate is within 0.03 of the cluster redshift provided by \citetalias{rines13}.

\begin{figure}
\begin{center}
\includegraphics[scale=0.35,trim = 0mm 10mm 0mm 15mm,clip]{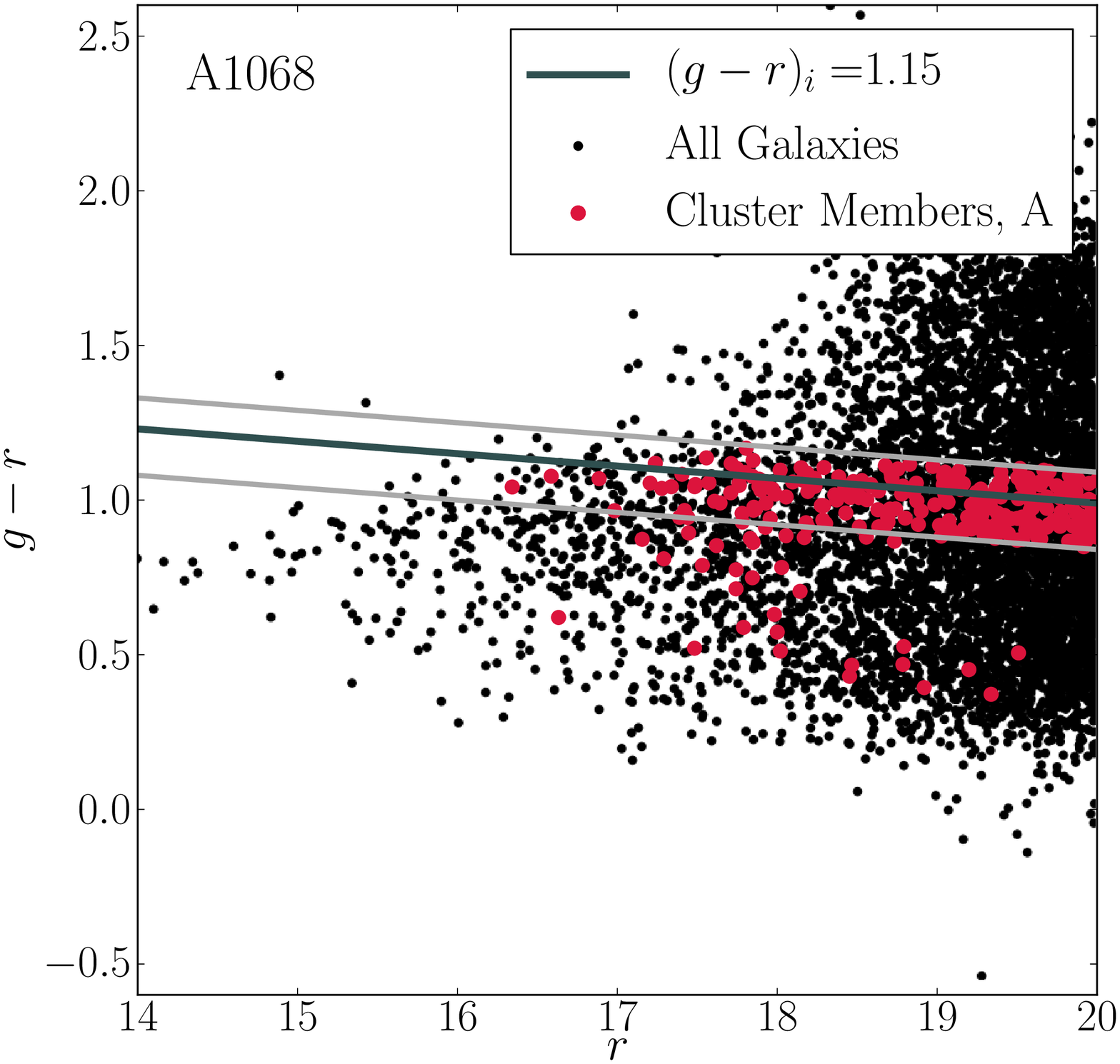}
\includegraphics[scale=0.35,trim = 0mm 10mm 0mm 15mm,clip]{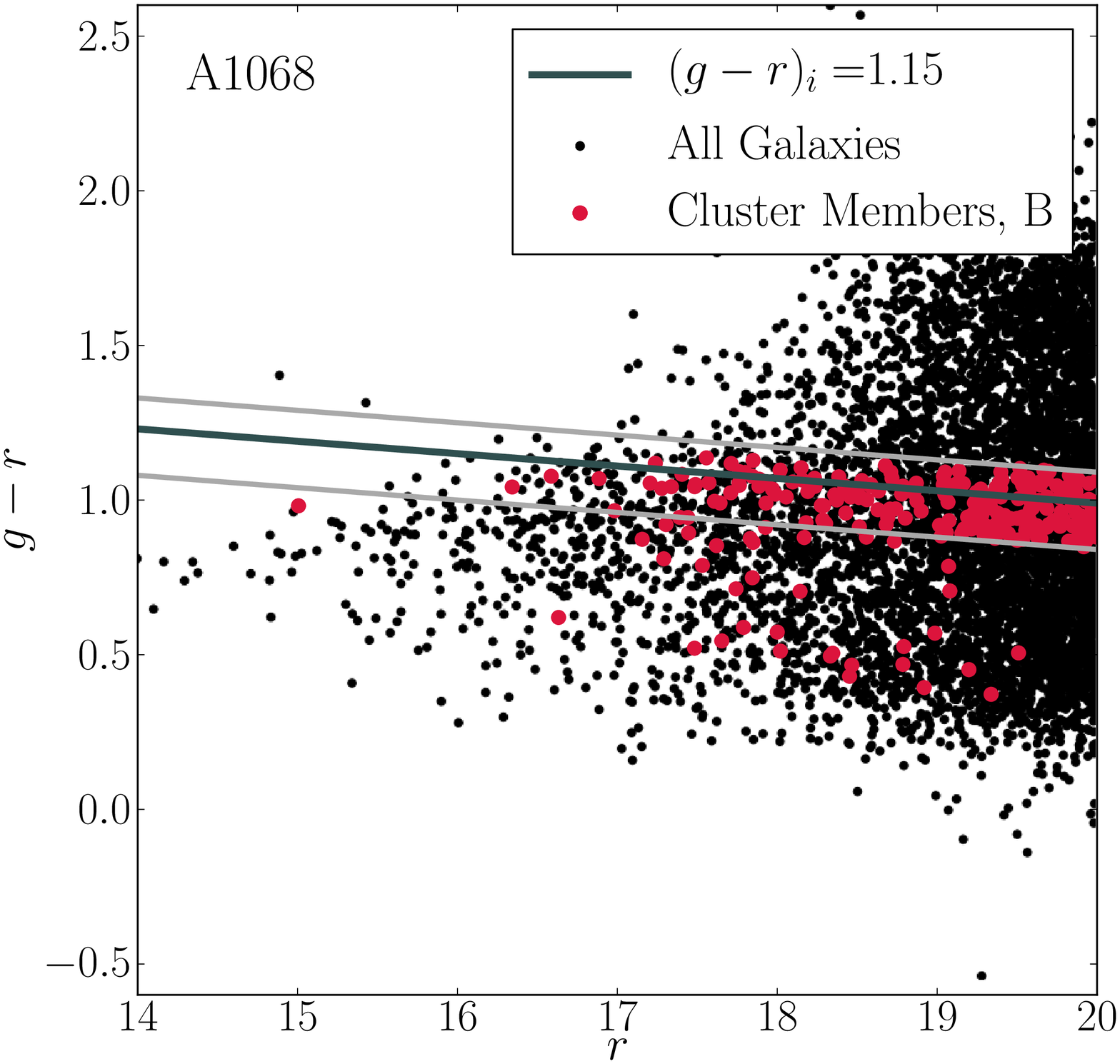}
\end{center}
\caption{\textit{Top:} The cluster member selection for example cluster A1068 using Selection A, which comprises galaxies selected using the red sequence method and galaxies with SDSS photometric redshifts, which add in some galaxies blueward of the red sequence. \textit{Bottom:} The cluster member selection for this same cluster using Selection B, which combines the red sequence, photometric redshifts, and spectroscopic redshifts.}\label{f:cm_pz}
\end{figure}

The population of galaxies with photometric redshift estimates of sufficient quality as indicated by the SDSS pipeline is too small for most clusters to use as the basis for our analysis; within the central $1.5\;h^{-1}\mathrm{Mpc}$, there are on average only about 60 galaxies with photometric redshifts satisfying $|z_c-z_g| < 0.03$ and subject to the above quality cuts, many of which were already picked up via the red sequence method. Additionally, basing a selection on these redshifts would exclude most cluster members with $r\gtrsim19$. Accordingly, as noted above, we must combine them with the galaxies selected via the red sequence method. This final selection, denoted Selection A, for an example cluster is shown in the lefthand panel of Figure~\ref{f:cm_pz}. 

\subsubsection{Additional Tests of the Cluster Member Selection}\label{s:addcm}
While neither the SDSS photometric redshifts nor the spectroscopy of \citetalias{rines13} identify enough cluster members for our analysis, we can use this additional information to refine the red sequence selection in a second selection. As before, after selecting red sequence cluster members using the procedure outlined in Section \ref{s:rs}, we add in bluer cluster members based on SDSS photometric redshifts as described in Section \ref{s:blue}. However, in this case, we can further refine the selection by also rejecting cluster members if the SDSS photometric redshifts are sufficiently secure and the members in question have redshifts in the range $|z_c-z_g| < 0.03$, where $z_c$ is the cluster redshift and $z_g$ is the photometric redshift estimate.

We can use the spectroscopic data of \citetalias{rines13} similarly. After matching their tables to the SDSS catalogs, we add in cluster members as identified by the spectroscopy and reject galaxies that were originally selected by the red sequence method or photometric redshifts but are identified as non-members by \citetalias{rines13}. This method of cluster member selection, which we term `Selection B,' is summarized in the righthand panel of Figure \ref{f:cm_pz}. This selection is more observationally expensive than Selection A, so it is worth testing both methods to see whether the additional information makes a difference in the results. On average, the numbers of cluster members selected by these two methods differ by about $5\%$ within $1.5\;h^{-1}\mathrm{Mpc}$ of the cluster center. We will test the effect of this selection on our analysis in Section~\ref{s:results}.

\begin{figure*}
\begin{center}
\includegraphics[scale=0.35]{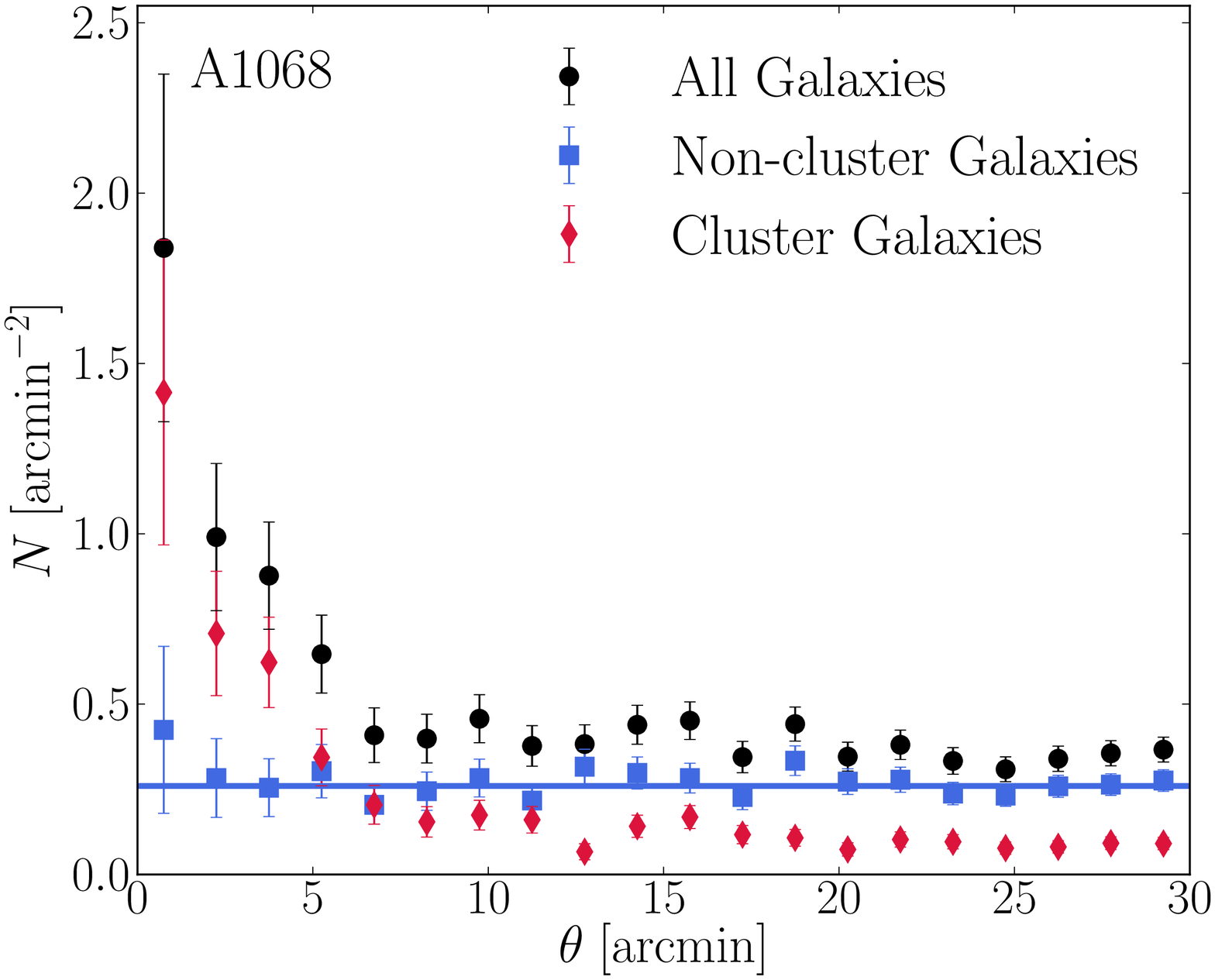}
\includegraphics[scale=0.35]{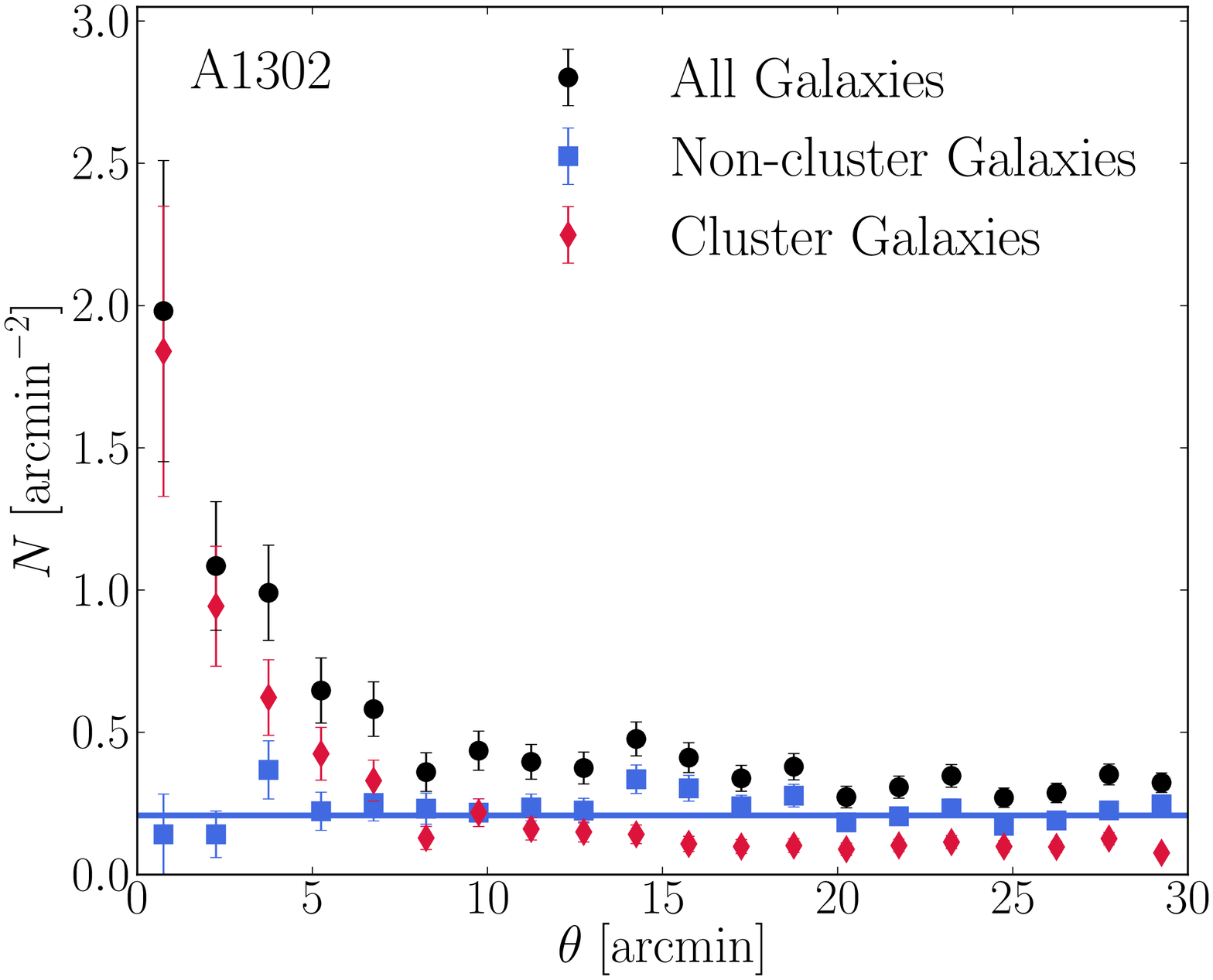}
\includegraphics[scale=0.35]{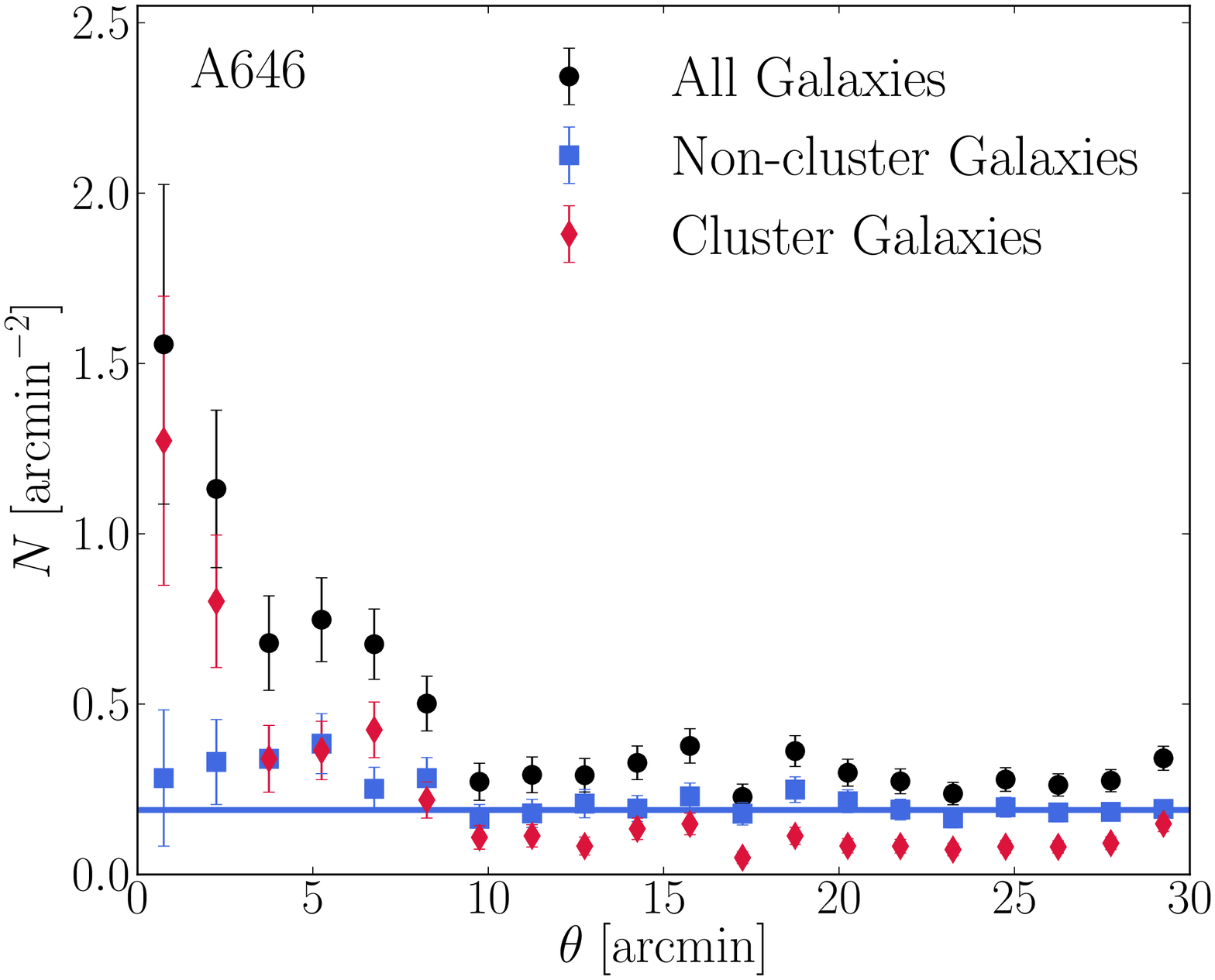}
\includegraphics[scale=0.35]{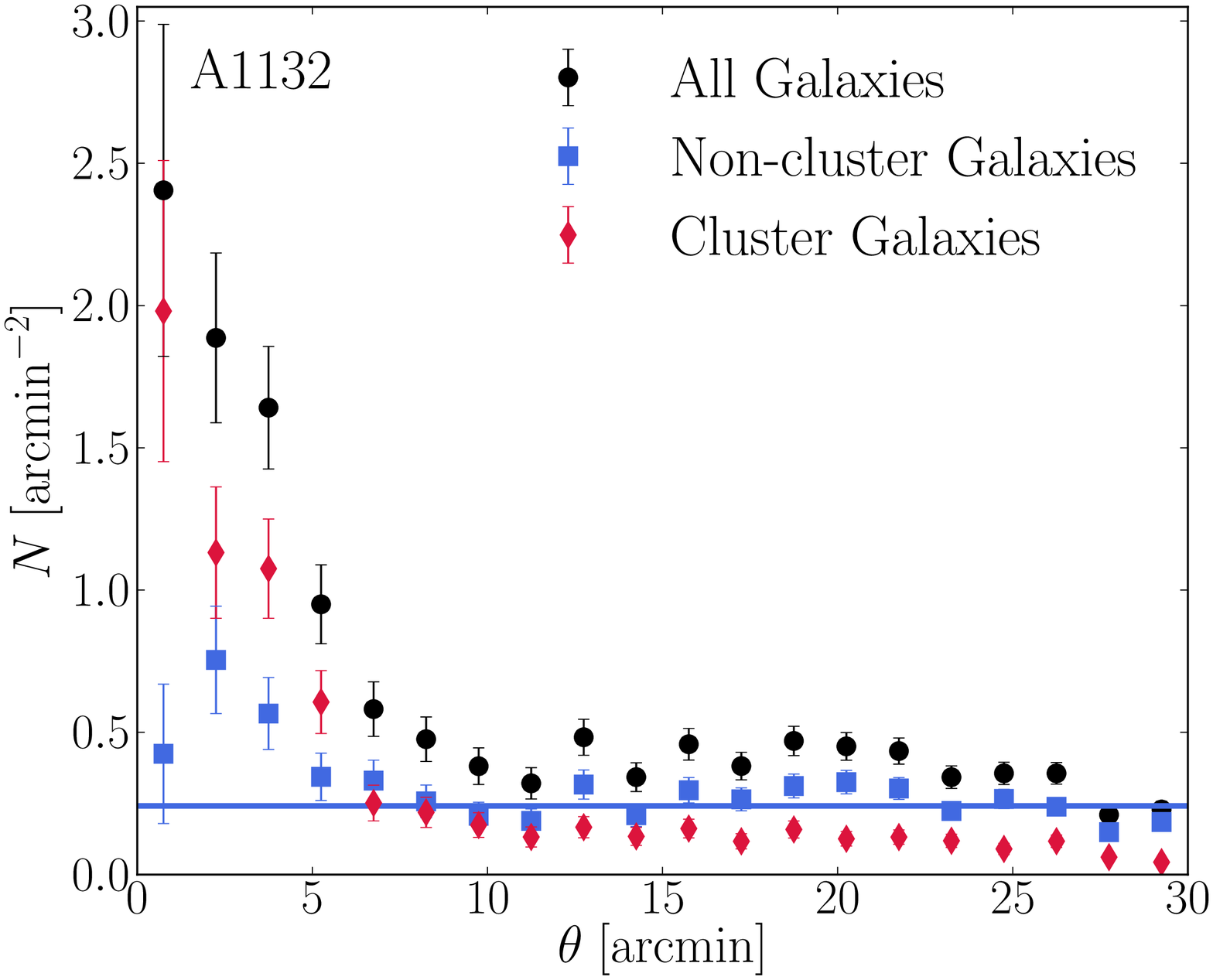}
\end{center}
\caption{Plots of the projected densities of galaxies with $r < 20$ in four cluster fields: A1068, A1302, A646, and A1132. Both the total galaxy density (black points) and the cluster density (red points) are expected to rise steeply towards the cluster center. The blue points show the density after removing cluster members from the total count, with a line drawn to indicate the average value beyond $20$ arcminutes for comparison. The non-cluster galaxy distribution should be roughly flat if we have adequately selected cluster members, as for A1068 and A1302. A646 and A1132, on the other hand, show some residual overdensity.}\label{f:ovc}
\end{figure*}

As an additional step in verifying our cluster member selection, we construct a projected radial density plot consisting of the total radial density profiles of all sources in our catalog (subject to $r < 20$) as well as the non-cluster density profile, computed by subtracting the counts of cluster members from the total number of galaxies in each bin. In the total profile, we expect to see a steep increase in galaxies at small radial distances that flattens out at large $R$. After subtracting out the contribution from these cluster members, the resulting non-cluster member profile should be roughly flat. Figure \ref{f:ovc} shows these density plots for four example clusters, using Selection A. 

\section{Method}\label{s:method}
\subsection{Approach}\label{s:approach}
Our fiducial analysis relies on cluster members selected as described in Section \ref{s:data}, subject to a magnitude cut of $r < 20$. We determine the number counts of galaxies in bins of $0.1\;h^{-1}$Mpc, from which we construct the projected number density profile of member galaxies, $N(R)$, with $R$ indicating the projected radius, using the cluster centers listed in \citetalias{rines13}. We use Poissonian error bars.

We employ two methods to test for evidence of the density jump. In the first, we fit the profiles using the fitting formula provided by \citetalias{diemer14} to incorporate their predicted density steepening:
\begin{align}\label{e:dk14}
\begin{split}
n_{\mathrm{DK}}(r) = n_\mathrm{in}(r)&\left[1+\left(\frac{r}{r_t}\right)^{\beta}\right]^{-\frac{\gamma}{\beta}}\\
&+n_m\left[b_e\left(\frac{r}{5R_{200}}\right)^{-s_e}+1\right].
\end{split}
\end{align}
As suggested by \citetalias{diemer14}, we fix $\beta = 6$ and $\gamma=4$, a choice that yields a dependence of $r_t$ on the mass accretion rate, $\Gamma$:
\begin{align}
r_t = \left(0.62+1.18e^{-2\Gamma/3}\right)\times R_{200}
\end{align}
We further select the NFW profile with two parameters, $n_s$ and $r_s$, as our inner density profile $n_\mathrm{in}$:
\begin{align}\label{e:nfw}
n_{\mathrm{NFW}}(r) = \frac{n_s}{r/r_s\left(1+r/r_s\right)^2}.
\end{align}
We note that \citetalias{diemer14} used the Einasto function \citep{einasto65} instead of the NFW for the inner profile, but in the regime of interest, the distinction between the two is negligible. We will refer to the model given by Equation~\ref{e:dk14} with $n_\mathrm{in}(r)=n_{\mathrm{NFW}}(r)$ as the `Density Jump' model, or `DJ' model in abbreviation.

Since the density jump feature occurs around the virial radius, we use a fitting range of $R<2R_{200}$, which is beyond the range that is typically fitted well by an NFW profile (we also exclude the inner $R < 0.1R_\mathrm{vir}$ in the fit, consistent with \citetalias{diemer14}). Accordingly, the two models that we will compare via fitting will be the DJ profile and a profile given by an NFW profile with an outer term:
\begin{align}\label{e:nfwo}
n(r) = n_{\mathrm{NFW}}(r)+n_m\left[b_e\left(\frac{r}{5R_{200}}\right)^{-s_e}+1\right].
\end{align}
These profiles are projected numerically and fitted to our galaxy density data. The free parameters in our fit are $n_s$, $r_s$, $b_e$, and $s_e$ for both profiles; the full DJ formula has one additional free parameter, $\Gamma$. We restrict the fits to the range $0 < \Gamma < 5$ and to reasonable values of $n_s$ and $r_s$, the latter of which is restricted by the value of the NFW concentration parameter, $c=R_{200c}/r_s$. Observations have suggested that this value is lower for galaxy profiles than for dark matter \citep[e.g.,][]{lin04,hansen05,budzynski12}, and we set the range as $2.0<c<6.0$ in our fits. We refer to the Appendix of \citetalias{diemer14} for appropriate ranges for $s_e$ ($0.5-2.0$) and $b_e$ ($0.1-4.0$). We additionally fix $n_m$, which will be discussed further in Section~\ref{s:radiusdefs}, and we select as our upper limit of integration $R = 10R_\mathrm{vir}$, same as used by \citetalias{diemer14}.

We then examine the results of the fits using the Akaike Information Criterion \citep[AIC;][]{akaike74} and the Bayesian Information Criterion \citep[BIC;][]{schwarz78}, which provide a means of comparing models fitted to data. As the use of these methods in astrophysics and cosmology has been discussed in a number of papers \citep[e.g.,][]{takeuchi00,liddle07,broderick11,tan12}, we simply mention the most salient qualities here and refer the reader to these works for further details. To apply these criteria, we compute the following statistics for each fit:
\begin{align}\label{e:aic_bic}
\mathrm{AIC} &= \chi^2+ 2p+\frac{2p(p+1)}{\mathcal{N}-p-1},\\
\mathrm{BIC} &= \chi^2+ p\ln(\mathcal{N}),
\end{align}
where $p$ is the number of parameters in the fit, $\mathcal{N}$ is the number of data points being fitted, and $\chi^2$ is the standard minimized goodness-of-fit parameter. In the case of the AIC, we have also included a correction term of $2p(p+1)/(\mathcal{N}-p-1)$ which is recommended for small values of $\mathcal{N}$ \citep{burnham02,burnham04}. The model that is preferred by these criteria is the one with the lower IC = AIC, BIC value. If we compute $\Delta\mathrm{IC} = \mathrm{IC}_{\mathrm{high}}-\mathrm{IC}_{\mathrm{low}}$, then, roughly, values of $\Delta\mathrm{IC} = 1-5$ are indicative of `positive' evidence in favor of the model with lower IC and values of $\Delta\mathrm{IC} > 5$ denote `strong' evidence \citep[e.g.,][]{liddle07,broderick11}.

Our second method consists of smoothing the profiles by fitting a smoothing cubic spline to our data. For a data set with measured values $y_i$ and errors $\sigma_i$ at a set of points $r_i$, the smoothing spline $f(r_i)$ is constructed to satisfy the condition
\begin{align}
\sum_{i=1}^{N}\left(\frac{y_i-f(r_i)}{\sigma_i}\right)^2 \le S,
\end{align}
where $S$ is a constant that interpolates between smoothing and fitting: that is, when $S=0$, the spline is forced to pass through every data point, so that there is no smoothing, whereas as $S$ is increased, the curve becomes smoother at the expense of the fit \citep{deboor01}. \citet{reinsch67} argues that the smoothing parameter $S$ should be chosen in the range $\mathcal{N}-\sqrt{2 \mathcal{N}}\le S \le \mathcal{N}+\sqrt{2 \mathcal{N}}$, where $\mathcal{N}$ is the number of data points over which we construct the spline, if the $\sigma_i$ are estimates of the standard deviation in $y_i$. We accordingly choose three values within this range to compare to an NFW fit: $S = \mathcal{N}-\sqrt{2 \mathcal{N}},\; \mathcal{N}, \;\mathcal{N}+\sqrt{2 \mathcal{N}}$. The NFW model is expected to be a good fit to the inner parts of the profile, so we use the analytical expression for the projected NFW density \citep[e.g.,][]{wright00} to fit the cluster galaxy density profiles within $R_{200}$, establishing the values of $n_s$ and $r_s$. We then calculate the logarithmic derivative $d\log(N)/d\log(R)$ of the splines for each cluster to test for the presence of the density jump feature, comparing it to the logarithmic derivative of the NFW fit.

\subsection{Fixed Parameters}\label{s:radiusdefs}
As noted in Section~\ref{s:intro}, we define measures of cluster size such that the mean mass density inside the radius $R_{\Delta}$ is $\bar{\rho} = \Delta\rho_b(z)$; commonly used values are $R_{500}$ and $R_{200}$. The other quantity that needs to be specified is the background density $\rho_b(z)$; one choice often used in observational work is the critical density $\rho_c(z) \equiv 3H(z)^2/8\pi G$, where $H(z)$ is the Hubble constant at redshift $z$, which is given by $H(z)^2=H_0^2\left[\Omega_m(1+z)^3+\Omega_{\Lambda}\right]$ with $H_0 = 100 \;h\; \mathrm{km/s/Mpc}$. Another choice is to use the mean matter density $\rho_m(z) = \rho_c(z)\Omega_m(z)$, where $\Omega_m(z) = \Omega_m(1+z)^3/\left[\Omega_m(1+z)^3+\Omega_{\Lambda}\right]$, which is used with $\Delta=200$.

As noted above, \citetalias{diemer14} use the mean matter density $\rho_m(z)$ to define the radius $R_{200}=R_{200m}$ used in the fitting formula given by Equation (\ref{e:dk14}). However, \citetalias{rines13} measures $R_{200}=R_{200c}$ for their sample of clusters using the critical density as reference, so we need to convert their measure to that of \citetalias{diemer14}. To do so, we note that a given mean density $\bar{\rho}$ may be written in two ways:
\begin{align}\label{eq:bar_rho}
\bar{\rho} = \Delta_c \rho_c(z) = \Delta_m \rho_m(z).
\end{align}
If we specify that $\rho = \rho_{\mathrm{NFW}}$, then in the outskirts of the cluster (i.e., including near $R_{200}$), we have:
\begin{align}
\bar{\rho}(R_{\Delta}) \propto \frac{1}{R_{\Delta}^3}.
\end{align}
Combining this relation with Equation~(\ref{eq:bar_rho}) yields
\begin{align}
\frac{R_{200m}}{R_{200c}} &= \left(\frac{1}{\Omega_m(z)}\right)^{1/3}.
\end{align}
For $z = 0.1-0.3$, this implies that
\begin{align}\label{e:r200convert}
R_{200m} \approx 1.4\times R_{200c}.
\end{align}
Accordingly, for simplicity, we will henceforth refer to $R_{200m}$ as $R_{200}$, and the values from \citetalias{rines13} will be converted to this measure using Equation (\ref{e:r200convert}).

Lastly, we need to establish an appropriate value for $n_m$. We note that the original prescription of \citetalias{diemer14} defines Equation (\ref{e:dk14}) in terms of mass densities $\rho$ rather than number densities $n$, and their fitting results assume a fixed value of $\rho_m(z) = \rho_c(z)\Omega_m(z)$. The translation into a number density needs to take into account the impact of the primary selection function by which we obtain cluster members, the red sequence method. In the absence of a cluster, this method would select a population of galaxies that lies in the appropriate region of color-magnitude space; in this case, the projected density of these galaxies is expected to be roughly constant across the field of view.

Recalling the outer profile term of Equation~(\ref{e:dk14}), 
\begin{align}
n_{\mathrm{out}}(r) = n_m\left[b_e\left(\frac{r}{5R_{200}}\right)^{-s_e}+1\right],
\end{align}
we can analytically determine the contribution to the surface density of the last, constant term. The surface density is the line-of-sight integral,
\begin{align}
N(R)= 2\int_R^{\infty}\frac{n(r)r}{\sqrt{r^2-R^2}}dr;
\end{align}
this integral diverges for a constant $n(r)$. However, in practice we must truncate this integral at some maximum radius $R_{\mathrm{max}}$. As noted above, \citetalias{diemer14} use $R_{\mathrm{max}} = 10 R_{\mathrm{vir}} \approx 9R_{200}$. In that case, 
\begin{align}
N_m(R) &=  2n_m\int_R^{9R_{200}}\frac{r}{\sqrt{r^2-R^2}}dr,\\
N_m(R) &= 2n_mR_{200}\sqrt{81-\left(\frac{R}{R_{200}}\right)^2}.
\end{align}
For the scales of interest in our fits, $R \lesssim 2R_{200}$ (and even a bit beyond), this value is roughly constant,
\begin{align}\label{e:Nmofnm}
N_m \approx 17.6 n_mR_{200}.
\end{align}

The redshift dependence of $N_m$ can be determined by applying the red sequence method in test fields that are not centered on low redshift clusters. If we select a population of galaxies with this method, then we can construct the projected density in a given radial bin $i$ as
\begin{align}
N_{m,i} &= \frac{\mathcal{N}_i}{\pi\left(R_{i,\mathrm{max}}^2-R_{i,\mathrm{min}}^2\right)},\\
&= \frac{1}{D_A(z)^2}\left[\frac{\mathcal{N}_i}{\pi\left(\theta_{i,\mathrm{max}}^2-\theta_{i,\mathrm{min}}^2\right)}\right],
\end{align}
where $\mathcal{N}_i$ is the number of galaxies in the $i$th bin and $D_A(z)$ denotes the angular diameter distance. We expect that the angular projected density (the term in brackets) over the radial range is a roughly constant value, which we denote by $\eta$. Then
\begin{align}
N_m(z) = \frac{\eta}{D_A(z)^2} = \frac{\eta(1+z)^2}{D_c(z)^2},
\end{align}
where $D_c$ is the comoving distance, which at the small redshifts considered here, is given by $D_c(z)\approx (c/H_0)z$. Upon absorbing the factor of $c/H_0$ into $\eta$, we have:
\begin{align}\label{e:Nmofz}
N_m(z) = \eta\frac{(1+z)^2}{z^2}.
\end{align}
To obtain the value of $\eta$, we apply our red sequence cuts in random test fields from SDSS (also subject to our initial magnitude cut of $r<20$)  and fit $N$ with the above function. This yields $\eta\approx0.08\;h^{2}\:\mathrm{Mpc}^{-2}$.

Accordingly, combining Equation (\ref{e:Nmofnm}) with (\ref{e:Nmofz}), we find that:
\begin{align}\label{e:nm}
n_m(z,R_{200}) = \frac{4.55\times10^{-3}}{R_{200}}\frac{(1+z)^2}{z^2}.
\end{align}
We fix this value individually for each cluster using its measured redshift and $R_{200}$ (the latter converted as discussed above) from \citetalias{rines13}.

\section{Results}\label{s:results}
\subsection{Fitting}\label{s:results_fitting}

The results of fitting the projected number density profiles with Equations (\ref{e:dk14}) and (\ref{e:nfwo}) are summarized in Table~\ref{t:ic_analysis} for both the fiducial method and variations, the latter of which will be discussed in the next section. However, as the DJ model has one additional parameter than the NFW+Outer model, we then compare the two fits using the AIC and BIC. We compute, in each case, $\Delta\mathrm{IC}=\mathrm{IC}_{\mathrm{NFW+O}}-\mathrm{IC}_{\mathrm{DJ}}$. Since lower values of IC are favored, this quantity will be positive if the criteria indicate evidence in favor of the DJ model. In Table~\ref{t:ic_analysis}, we thus first list in the column, `\% $\chi^2_{\mathrm{DJ}}/\mathrm{ndf}<\chi^2_{\mathrm{NFWO}}/\mathrm{ndf}$,'  the number of galaxy clusters for which the reduced $\chi^2$ is lower for the DJ model and second the number of galaxy clusters whose fits pass a general quality cut, including requiring that $\chi^2/\mathrm{ndf} < 3$ for the DJ model and a more generous $\chi^2/\mathrm{ndf} < 5$ for the other. The next two columns indicate how many clusters have $\Delta\mathrm{IC}=\mathrm{IC}_{\mathrm{NFW+O}}-\mathrm{IC}_{\mathrm{DJ}} > 0$ for each criterion, which would suggest evidence in favor of the DJ model. The last two columns indicates how many clusters have one of the $\Delta\mathrm{IC} > 5$, indicating strong evidence in favor of this model.

\begin{figure*}
\begin{center}
\includegraphics[scale=0.41]{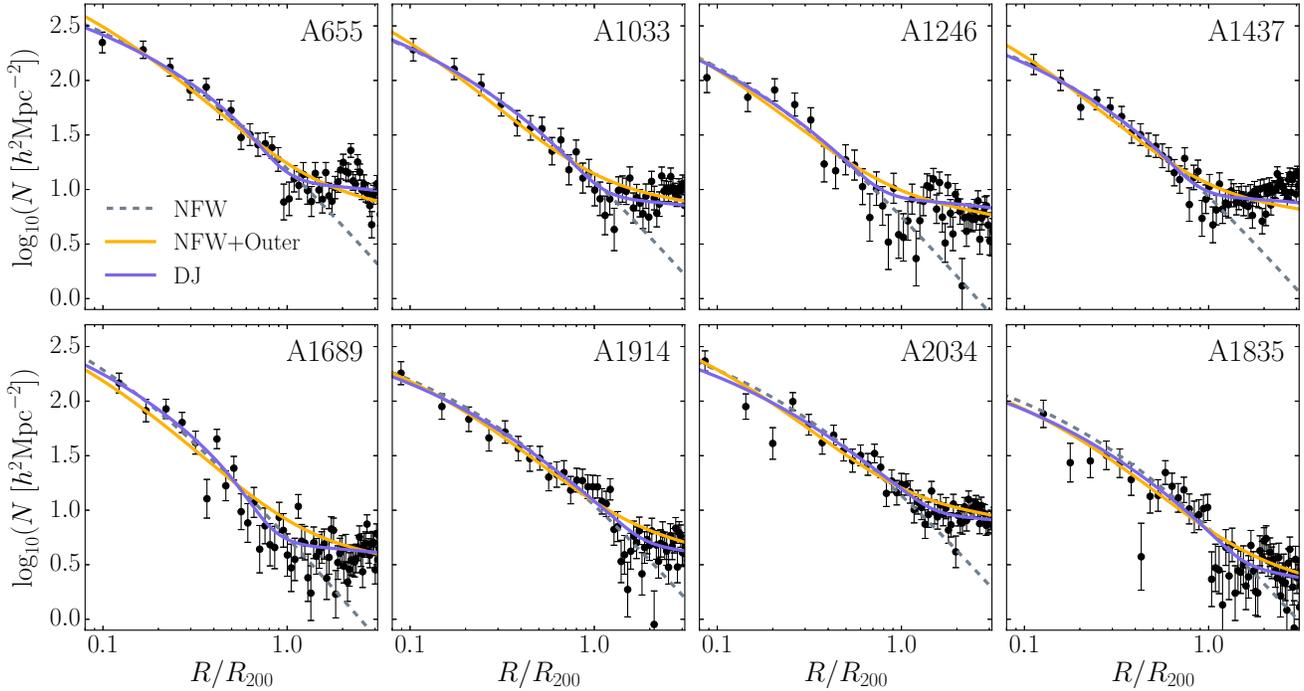}
\end{center}
\vspace{-0.1in}\caption{Plots of the projected number density profiles for the 8 clusters with the highest $\Delta\mathrm{AIC}$ and $\Delta\mathrm{BIC}$ values in the fiducial analysis using Selection A. Three fitted functions are shown: a base NFW, fitted interior to $R_{200}$, an NFW+Outer model, given by Equation (\ref{e:nfwo}) and fitted interior to $2R_{200}$, and the full Density Jump model given by Equation~\ref{e:dk14} with an inner NFW profile, fitted interior to $2R_{200}$.}\label{f:fits_fidA}
\end{figure*}

\begin{figure*}
\begin{center}
\includegraphics[scale=0.41]{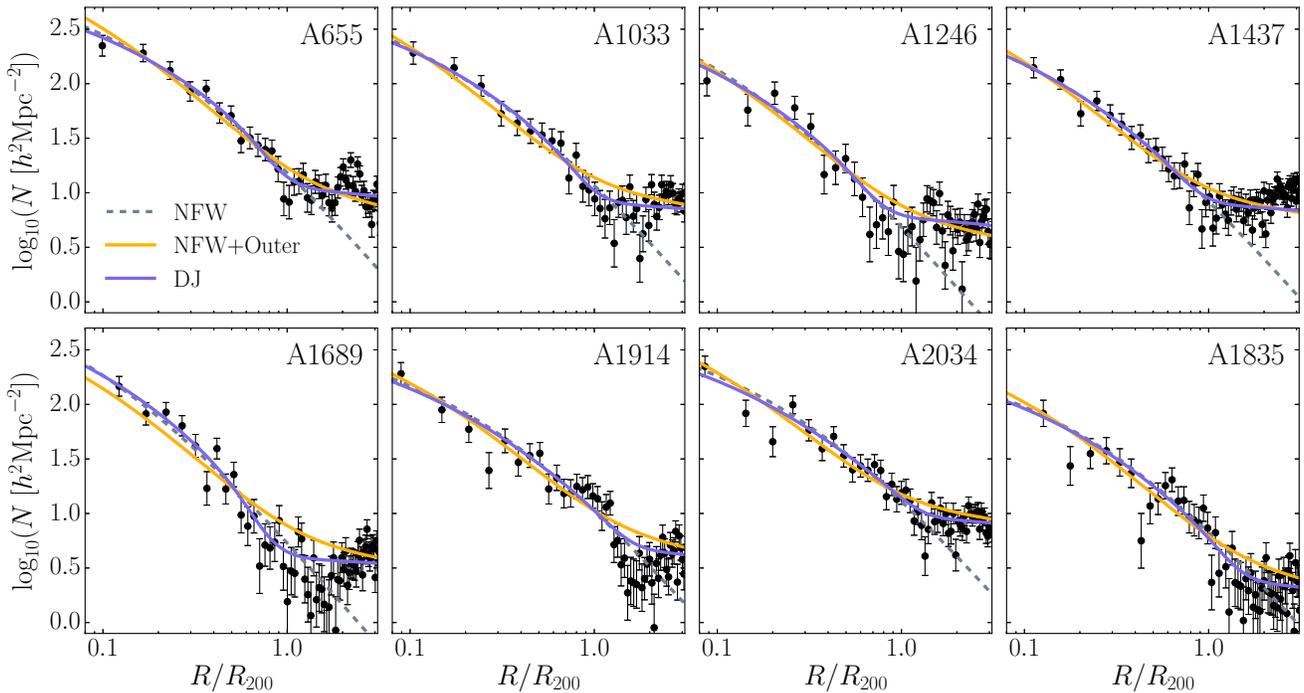}
\end{center}
\vspace{-0.1in}\caption{Plots of the projected number density profiles for the 8 clusters with $\Delta\mathrm{IC}>5$ using both selection methods, with fits shown for the profiles constructed using Selection B. The curves are the same as in Figure \ref{f:fits_fidA}.}\label{f:fits_fidB}\vspace{-0.1in}
\end{figure*}

Our fiducial method uses a binning of $\Delta R = 0.1\;h^{-1}$Mpc and employs a magnitude cut of $r<20$. We use two selection methods to identify cluster members; as discussed in more detail in Section \ref{s:data}, Selection A is based on SDSS photometry and Selection B refines the first with \citetalias{rines13} spectroscopy. We see that refining the cluster member selection yields a higher number of clusters for which the DJ model is a better description than a simple NFW model with an outer term. However, the results of individual clusters are consistent between the selection methods -- the seven clusters with the strongest evidence in favor of the density jump using Selection A (A655, A1033, A1246, A1437, A1689, A1914, and A2034) continue to have $\Delta\mathrm{IC}>5$ using the second selection. These clusters (along with the more marginal case A1835, which has only one $\Delta\mathrm{IC}>5$ using Selection A) are shown in Figure~\ref{f:fits_fidA} for Selection A fits and Figure~\ref{f:fits_fidB} for Selection B; however, using Selection B several other clusters also reveal strong (and for some even stronger) evidence for a density jump. In both cases, the cluster with the strongest evidence for a density jump is A1689.  

Furthermore, the increase in the number of clusters with IC evidence in favor of the DJ model using Selection B suggests that it may be beneficial to investigate these profiles further with a cleaner cluster member selection, with either a sample of galaxies with high-quality photometric redshift estimates or with a sufficiently dense sample of galaxies with secure spectroscopic redshifts. Additionally, a few of these clusters (particularly A1246) show some residual overdensity at roughly $\theta < 5'$ of the cluster center after the cluster member selection (c.f. Figure~\ref{f:ovc}), which could be indicative either of having missed some population of galaxies in our selection or that there are some other agglomerations of galaxies along the line of sight that contribute to the overdensity. High-quality, dense redshift estimates would help resolve this ambiguity.

We additionally test the fits by making the radial bins twice as large ($\Delta R = 0.2\;h^{-1}$Mpc). In this case, the reduced $\chi^2$ values are a bit worse for Selection A, as indicated in Table~\ref{t:ic_analysis}. We find that the larger binning yields the greatest drop in the number of clusters with evidence favorable towards the DJ model using the AIC; the BIC results, however, are virtually unchanged from the earlier case. The discrepancy is likely caused by the smaller number of data points over which we fit relative to the number of parameters, which makes affects the correction term for the AIC.

\begin{table*}
\renewcommand{\arraystretch}{1.2}
\addtolength{\tabcolsep}{4pt}
\caption{Information Criteria Results}
\centering
\label{t:ic_analysis}
\begin{tabular}{l c c c c c}
\hline \hline
Method & \% $\chi^2_{\mathrm{DJ}}/\mathrm{ndf}<\chi^2_{\mathrm{NFWO}}/\mathrm{ndf}$ & $\Delta\mathrm{AIC} > 0$ & $\Delta\mathrm{BIC}> 0$  & $\Delta\mathrm{AIC} > 5$ & $\Delta\mathrm{BIC}> 5$ \\ [0.5ex]
\hline
Fiducial, Sel. A & 38/56 & 18 & 22 & 8 & 7 \\
Fiducial, Sel. B & 43/56 & 32 & 32 & 16 & 20 \\
$\Delta R = 0.2\;h^{-1}$Mpc, Sel. A & 34/55 & 12 & 22 & 5 & 7  \\
\hline
\end{tabular}
\end{table*}

\subsection{Smoothing Splines}\label{s:spline}
In addition to fitting the projected galaxy density profiles with Equations (\ref{e:dk14}) and (\ref{e:nfwo}), we also fit a smoothing cubic spline as described in Section~\ref{s:approach}. This latter approach has the benefit of being model-independent; we use the spline fits to construct a smooth logarithmic derivative of the profile to test for the existence of a density jump.

\begin{figure}
\begin{center}
\includegraphics[scale=0.33,trim=2mm 0mm 0mm 0mm,clip]{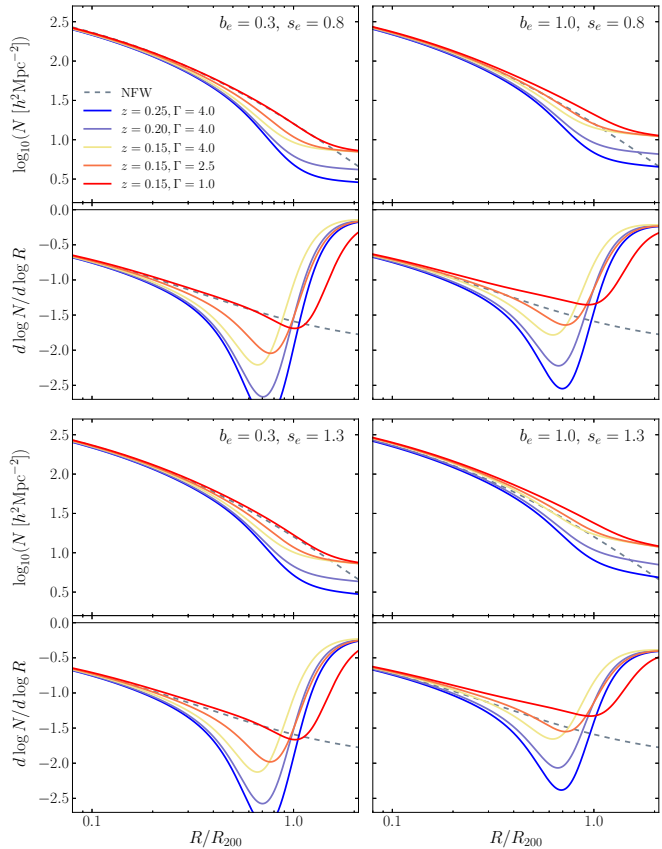}
\end{center}
\caption{The predicted signature of the density jump is a steepening of slope that can be analyzed using the logarithmic derivative of the \citetalias{diemer14} model for four combinations of $s_e$ and $b_e$ values. The redshift dependence enters via $n_m(z,R_{200})$ as given by Equation (\ref{e:nm}), and we show the result for three values of $z$ in the range of the \citetalias{rines13} clusters at fixed $\Gamma=4$. For the $z=0.15$ case we also show the variation of the profile with $\Gamma$. An NFW profile (dashed) is drawn for comparison. All other model parameters are fixed for all curves shown.}\label{f:dkder}
\end{figure}

First, to provide context for the spline results, in Figure~\ref{f:dkder}, we plot the DJ model at fixed $\Gamma=4$ for various redshifts, which determine $n_m$, and for an example fixed redshift $z=0.15$, we also show the variation with $\Gamma$. The bottom panel of each plot shows the corresponding logarithmic derivatives. The effect of the redshift dependence is to make the density jump steeper at higher redshift due to the lower background $n_m$. At a fixed $n_m$, a smaller value of $\Gamma$ means a shallower density jump, whose maximum amplitude slope is attained at larger radii, although the values of $b_e$ and $s_e$, which are free parameters in our fits, also contribute to the variation in the amplitude.

Figure~\ref{f:dkder} can be compared to Figure 13 of \citetalias{diemer14}. While the behavior of the density jump is qualitatively the same, we note that in our analysis, the maximum slope amplitudes are smaller than those predicted by \citetalias{diemer14} due to the elevated background; accordingly, in our case the maximum slope amplitudes are governed not only by the mass accretion rate $\Gamma$ but also the redshift of the cluster. However, the density jump location also depends on $\Gamma$. 

\begin{figure*}
\begin{center}
\includegraphics[scale=0.47]{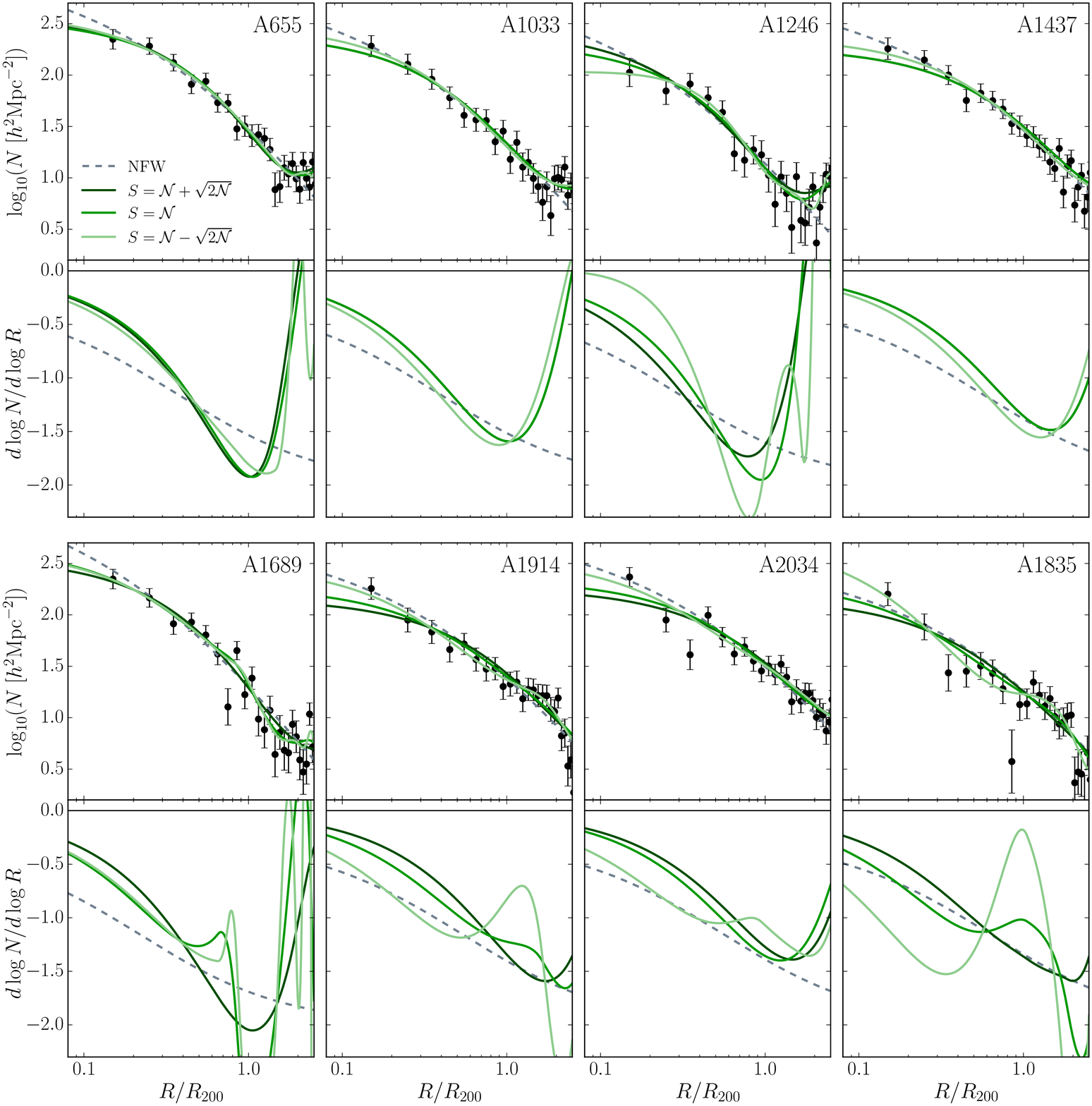}
\end{center}
\caption{The top panel shows spline fits for various values of the smoothing parameter $\mathcal{S}$ for the clusters in Figure~\ref{f:fits_fidA}. The lower panel displays the logarithmic derivatives of the splines. An NFW fit and its logarithmic derivative (dashed line) are shown for comparison.}\label{f:spline_fidA}
\end{figure*}

\begin{figure*}
\begin{center}
\includegraphics[scale=0.47]{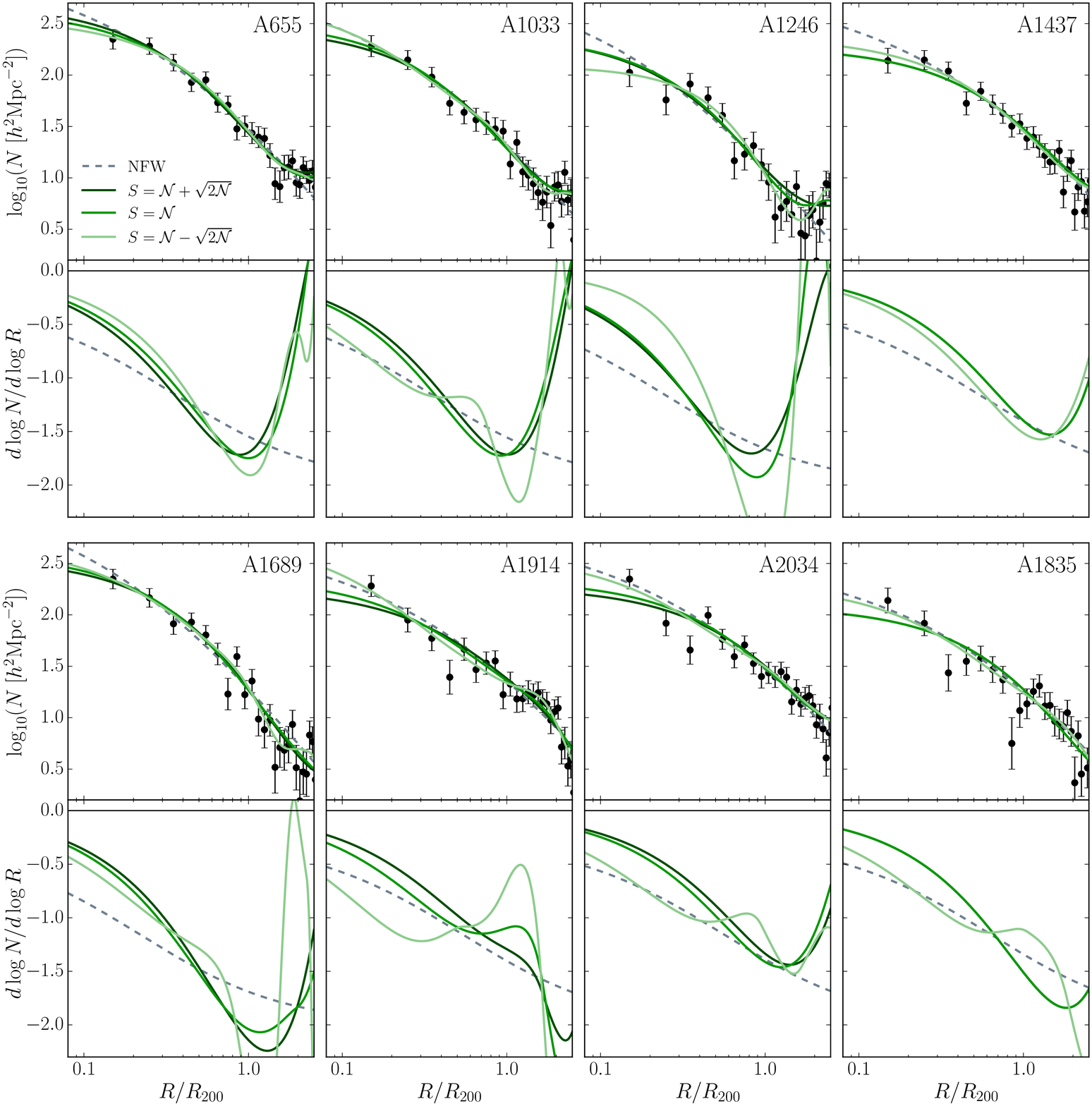}
\end{center}
\caption{Same as Figure~\ref{f:spline_fidA}, but for Selection B profiles.}\label{f:spline_fidB}
\end{figure*}

Accordingly, we model our spline fits on Figure~\ref{f:dkder}. We show the cubic spline fits for the same subset of clusters selected via the information criteria in the fiducial analysis. Figure~\ref{f:spline_fidA} shows the spline fits for Selection A, while Figure~\ref{f:spline_fidB} is its counterpart for Selection B. The spline fits do appear to pick up a modest steepening of slope, consistent with the low redshift of these clusters. Since at low redshift we can expect to almost exclusively detect very large density jumps ($\Gamma\approx5$), which are likely uncommon, it is perhaps not surprising that the number of clusters for which the information criteria suggest strong evidence for the jump is fairly low.

\section{Discussion}\label{s:discussion}
We have tested whether cluster galaxy density profiles show evidence for a density jump feature near the virial radius using two methods: profile fitting and spline smoothing. We have examined the evidence in favor of the presence of a steep density jump in the galaxy density profiles of clusters, and also investigated the results via spline smoothing.

There does appear to be some dependence of our results on cluster redshift. In our fiducial analysis of Section~\ref{s:results_fitting}, the seven clusters that showed strong evidence (both $\Delta\mathrm{IC}>5$) for the density jump using Selection A spanned the redshift range of roughly $z = 0.1-0.2$, which omits the higher redshift clusters (although an eighth cluster, A1835, with one $\Delta \mathrm{IC} > 5$ value, is at $z = 0.25$). However, employing Selection B enlarges the sample of strong evidence clusters, and extends this to about  $z = 0.1-0.25$, effectively the entire range of \citetalias{rines13}. More massive clusters  tend to show stronger evidence for a density jump: these same 8 clusters have virial masses in the range $M_{\mathrm{vir}} = (2-11)\times10^{14}\:h^{-1}M_{\odot}$, while a significant fraction of the clusters in the \citetalias{rines13} sample have $M_{\mathrm{vir}} = (0.3-2)\times10^{14}\:h^{-1}M_{\odot}$ (and may also be considered galaxy groups). This trend continues even when considering the additional clusters that pass the criteria using Selection B. This behavior can be expected, as higher mass clusters tend to have higher values of $\Gamma$ and their profiles are likely to be better sampled than lower mass systems.

We see variations based on the cluster member selection. Our primary selection uses only photometric data, but redoing the analysis with the inclusion of the \citetalias{rines13} spectroscopy does shift the results; in particular, the refined selection yields a larger sample of clusters that show some evidence for the jump. This suggests that additional data is required to make a firm detection of the density jump. This additional data could be in the form of dense spectroscopy out to large radii of cluster member galaxies, building upon the \citetalias{rines13} catalog, which would provide the most secure cluster member determination. Otherwise, high-quality photometric redshifts could also improve upon our estimates.

Additionally, it is worth noting that in interpreting these results it is necessary to keep in mind some of the other factors that can impact the jump signature. In particular, cluster asphericity and contamination by other small groups and clusters along the line of sight can contribute to a diminishing of the jump signal. The latter of these can be addressed with redshift estimates. For the first, we note that the signature of the jump becomes more pronounced and thus easier to detect when $\Gamma$ is large; however, clusters with large values of $\Gamma$ may be more likely to have disturbed shapes and substructures, which when examined in projection could obscure the signature of the density jump.

More generally, it would be useful to compare the results in Table~\ref{t:ic_analysis} to simulations. Mock halo catalogs could give an indication of the conditions needed for a cluster to show a discernible jump and thus the fraction of clusters in which we can expect to find this signature, which would provide context for the fractions we find using observational data. While this is a promising avenue for future analyses, such a comparison is beyond the scope of this work.

\section{Summary and Conclusions}
Using the cluster sample of \citetalias{rines13} and optical data from SDSS, we have searched for the signature of a density jump feature near the virial radius ($\sim R_{200}$) predicted by the simulations of \citetalias{diemer14}. Our fiducial analysis selects cluster members from SDSS photometric catalogs using the red sequence method and photometric redshifts, and compares this to a selection refined by the inclusion of the spectroscopy of \citetalias{rines13}. After constructing the radial density profiles of the clusters, we fit two models, one with a density jump --- Equation~(\ref{e:dk14}) --- and one without --- Equation~(\ref{e:nfwo}) --- and used the Akaike and Bayesian Information Criteria (AIC and BIC) to examine the evidence in favor of the density jump model. These criteria indicate that, using our fiducial methods, at least 10\% showed strong evidence ($\Delta\mathrm{IC}>5$) for the model that includes a density jump. The clusters with strong evidence for the density jump tend to have a higher mass ($M_{\mathrm{vir}} \gtrsim 2\times10^{14}h^{-1}M_{\odot}$), as expected. The cluster with the strongest evidence for the jump in this sample is Abell 1689.

We examined varying some of the parameters in our analysis. We find that using a larger bin size yields fewer clusters with strong evidence (if we stipulate a strong result for both criteria; the AIC appears to be most affected by the binning) for the density jump. However, if we require only one of the information criteria to yield strong evidence, then we are in excellent agreement with the fiducial analysis. 

We additionally tested our results using cubic smoothing splines. The spline analysis appears to indicate some steepening of slope in these clusters, but the comparison to the theoretical predictions of \citetalias{diemer14} is complicated by background and projection effects. In the case of the former, a high background can diminish the signal, while the latter acknowledges that since clusters can have diverse shapes, the density jump is not necessarily localized at a single radius, which can lead to modifications of the signal in projection. 

Finally, varying the cluster member selection methods does appear to have an effect on the results; in particular, including the most secure cluster member indicators (spectroscopic redshifts) increases the number of clusters for which we have strong evidence in favor of a model with a density jump. Thus, our conclusions are limited by the availability of data for the identification of galaxy cluster members. While the red sequence method provides a means of identifying the majority of cluster members, which tend to be red, early type galaxies, it does not readily provide a means of rejecting interloping red galaxies from higher redshifts or for including the smaller population of member galaxies blueward of the red sequence. While the spectroscopic observations of \citetalias{rines13} provide an excellent start, it would be useful to extend their samples with additional redshift estimates to improve the completeness in the cluster outskirts, mitigating observational selection effects that can mimic a density steepening.

Accordingly, at present, neither the extant photometry nor spectroscopy of these clusters provides adequate numbers of confirmed member galaxies out to sufficiently large radii to significantly improve upon our methods. However, in order to better constrain the density jump feature and precisely determine its amplitude and position, future efforts will require either secure photometric redshifts or additional spectroscopy to refine the cluster galaxy density profiles. Future work using mock halo catalogs could help pinpoint the conditions under which clusters can be expected to show a discernible density jump to help guide these observational endeavors.

\acknowledgments
We are very grateful to Andrey Kravtsov and Benedikt Diemer for their detailed and helpful comments on a draft of this paper. We would also like to thank the staff of helpdesk@sdss.org for advice regarding downloading data from SDSS. 

This work was partly supported by the National Science Foundation Graduate Research Fellowship under Grant No. DGE-1144152. This work was also supported by NSF Grant No. AST-1312034. This work relied on tools from Numpy \citep{numpy11}, Scipy \citep{oliphant07}, and Matplotlib \citep{hunter07}, as well as ROOT (http://root.cern.ch/).

This paper relies on data from SDSS-III. Funding for SDSS-III has been provided by the Alfred P. Sloan Foundation, the Participating Institutions, the National Science Foundation, and the U.S. Department of Energy Office of Science. The SDSS-III web site is http://www.sdss3.org/.

SDSS-III is managed by the Astrophysical Research Consortium for the Participating Institutions of the SDSS-III Collaboration including the University of Arizona, the Brazilian Participation Group, Brookhaven National Laboratory, Carnegie Mellon University, University of Florida, the French Participation Group, the German Participation Group, Harvard University, the Instituto de Astrofisica de Canarias, the Michigan State/Notre Dame/JINA Participation Group, Johns Hopkins University, Lawrence Berkeley National Laboratory, Max Planck Institute for Astrophysics, Max Planck Institute for Extraterrestrial Physics, New Mexico State University, New York University, Ohio State University, Pennsylvania State University, University of Portsmouth, Princeton University, the Spanish Participation Group, University of Tokyo, University of Utah, Vanderbilt University, University of Virginia, University of Washington, and Yale University.

\end{document}